**Big Bang Nucleosynthesis and Primordial Black Holes**

C. Sivaram and Kenath Arun
Indian Institute of Astrophysics, Bangalore

**Abstract:** There are ongoing efforts in detecting Hawking radiation from primordial black holes (PBH) formed during the early universe. Here we put an upper limit on the PBH number density that could have been formed prior to the big bang nucleosynthesis era, based on the constraint that the PBH evaporation energy consisting of high energy radiation not affect the observed abundances' of elements, by disintegrating the nuclei. This puts a limit of $\Omega_{PBH} \leq 10^{-15}$

The light elements in the universe were made in the big bang nucleosynthesis (BBN) in the first 3 minutes when temperature was ~$10^9$ K. The synthesis of these elements as the hot universe expands is predicted by the theory which matches with observations [1]. That is 0.245 abundance for hydrogen, ~$10^{-5}$ of deuterium and $He^3$ and ~$10^{-9}$ of $Li^7$. Beyond three minutes, the temperature became too low for the synthesis of any higher elements beyond $Li^7$. [2]

The total number of nucleons in the universe is conserved and is $N_p \approx 10^{78}$, then the number of deuterium nuclei produced in the BBN is given by:

$$n_D \approx 10^{78} \times 10^{-5} \approx 10^{73} \qquad \text{... (1)}$$

The binding energy of the deuterium nucleus is of the order of 2MeV (~$10^{-6}$ ergs). The total energy required to disintegrate all the deuterium nuclei is:

$$E_D \sim 10^{-6} ergs \times 10^{73} = 10^{67} ergs \qquad \text{... (2)}$$

We consider the possibility that the PBH produced at earlier epochs than the BBN era, and having lifetime ~100s (comparable to the duration of the BBN) can act as a source of high energy gamma ray which could disintegrate the deuterium nuclei [3]. The mass of PBH with lifetime of ~100s is ~$10^{10}$g and the energy released in high energy radiation is: [4]

$$M_{BH} c^2 \approx 10^{31} ergs \qquad \text{... (3)}$$

For the required $10^{67}$ ergs, to disintegrate all the deuterium nuclei, we need $10^{36}$ PBHs each of mass ~$10^{10}$g. PBHs of mass less than this would have already evaporated and those with higher masses will have a lower rate of evaporation and will not contribute much to the energy. The total mass of all these PBH will be ~$10^{46}$g. The total baryonic mass in the universe (which remains conserved) is ~$10^{55}$g; this implies that the Ω factor for the PBH is about:

$$\Omega_{PBH} \approx 10^{-10} \qquad \text{... (4)}$$

However this is only in comparison to the baryonic matter. The total energy (mass) of the universe at the era of the BBN is

$$E_{BBN} \approx aT_R^4 \times \text{volume of universe at BBN} \qquad \text{... (5)}$$

Using $RT =$ constant, (present $T_0 \approx 3K$; $R_0 \approx 10^{28}\,cm$), gives a total energy of

$$E_{BBN} \approx 10^{88} \times 10^{-6}\,ergs \approx 10^{82}\,ergs \qquad \text{... (6)}$$

Where, the number of photons in the radiation background is conserved and is $\sim 10^{88}$ and the energy of a photon at BBN is $\sim 10^{-6}\,ergs$. And since at early epoch, $\Omega = 1$, we have:

$$\Omega_{PBH} = \frac{10^{67}\,ergs}{10^{82}\,ergs} \approx 10^{-15} \qquad \text{... (7)}$$

The much heavier PBH masses which survive to the present would be expected to be produced with smaller abundances, so their $\Omega$ is likely to be $<< 10^{-15}$, at present.

This is consistent with any non-observation of hawking radiation from PBH's so far! [5]

***Note:*** $\Omega_{PBH} \leq 10^{-15}$ is really an upper limit as we assumed all the deuterium nuclei is disintegrated!